\documentclass[useAMS,usenatbib]{mn2e}
\usepackage{float,graphics,epsfig,array,amsmath,amssymb}
\usepackage{times}
\usepackage{color}
\usepackage{natbib}
\def\araa{ARA\&A}%
\def\apj{ApJ}%
\def\apjs{ApJS}%
\def\aap{A\&A}%
\def\mnras{MNRAS}%
\def\pasj{PASJ}%
\def\physrep{Phys.~Rep.}%
\newcommand\ion[2]{#1$\;${\scshape{#2}}}
\newcommand{\ltsima}{$\; \buildrel < \over \sim \;$}
\newcommand{\simlt}{\lower.5ex\hbox{\ltsima}} 
\newcommand{\gtsima}{$\; \buildrel > \over \sim \;$}
\newcommand{\simgt}{\lower.5ex\hbox{\gtsima}} 

\newcommand{\fexxv}{\ensuremath{\mbox{\ion{Fe}{xxv}}}}
\newcommand{\xmm}{{XMM-\emph{Newton}}}

\newcommand{\lum}{erg~s$^{-1}$}
\newcommand{\flux}{{erg~cm$^{-2}$~s$^{-1}$ }}
\newcommand{\nh }{cm$^{-2}$}
\newcommand{\nhsym}{$N_{\rm{H}}$}

\newcommand{\chandra}{{\emph{Chandra}}}

\newcommand{\suzaku}{{\emph{Suzaku}}}

\newcommand{\swift}{{\emph{Swift}}}
\newcommand{\logxi}{erg cm s$^{-1}$}

\title[ NGC454: unveiling a new ``changing look'' AGN  ]{  NGC454: unveiling a new ``changing look'' AGN  }
\author[Marchese et al.]{E. Marchese$^{1,}$ $^{2}$\thanks{E-mail: elena.marchese@brera.inaf.it}, V. Braito$^{1,}$ $^{3}$\thanks{E-mail: valentina.braito@brera.inaf.it}, R. Della Ceca$^{1}$, A. Caccianiga$^{1}$,  P. Severgnini$^{1}$\\
 $^{1}$INAF-Osservatorio Astronomico di Brera, via Brera 28, 20121 Milano, Italy\\
$^{2}$Universit\`a degli studi di Milano-Bicocca, Piazza dell'Ateneo Nuovo, 1 - 20126, Milano \\
$^{3}$X-Ray Astronomy Observational Group, Department of Physics and Astronomy, Leicester University, Leicester LE1 7RH, UK\\
}

\begin{document}

\date{}

\pagerange{\pageref{firstpage}--\pageref{lastpage}} 

\maketitle

\label{firstpage}

\begin{abstract}

 We present a detailed analysis of  the  X-ray spectrum  of  the Seyfert 2
galaxy NGC454E, belonging to the interacting system NGC454.  Observations
performed    with \suzaku, \xmm \ and \swift\ allowed us to detect a dramatic
change in the curvature of the 2--10 keV spectrum, revealing a significant  
variation of the absorbing column density along the line of sight (from $\sim 1
\times 10^{24}$\nh\   to $\sim 1\times 10^{23}$\nh). Consequently, we propose
this source as a new member of the class of ``changing look'' AGN, i.e. AGN that
have been observed  both in Compton-thin  (\nhsym =$10^{23} \rm cm^{-2}$) and
reflection dominated states (Compton-thick, \nhsym $>10^{24} \rm cm^{-2}$). Due to the quite long
time lag (6 months) between the \suzaku\ and \xmm\ observations we cannot infer
the possible location of the obscuring material causing the observed
variability. In the  6--7 keV range the
\xmm\ observation also shows a clear signature  of
the presence of an ionized absorber. Since this feature is not detected during the \suzaku\ observation (despite its detectability), the simplest interpretation is that the ionized absorber is also variable; its location is estimated to be within $\sim 10^{-3} $ pc from the central black hole, probably much closer in than the  rather neutral absorber.     

\end{abstract}


\section{Introduction}
 
There is now a general consensus that Active Galactic Nuclei (AGN) are powered
by accretion of matter onto a supermassive black hole (SMBH), located at center
of almost all massive galaxies.   It is also clear that, according to  the
Unified Model of AGN (\citealt{Antonucci93}), the difference between type 1 and
type 2 AGN can be explained  through orientation effects between our line of
sight to the nucleus and ``circum-nuclear material".   However, the geometry,
size and physical state of this circum-nuclear matter are still a matter of
debate. In particular, the AGN X-ray spectra are complex and  consist of  
multiple components (see \citealt{Turner09}, \citealt{Done2010} for a review), which are  all
intimately related to the  still poorly  understood condition of the matter near
the nucleus. This circum-nuclear gas imprints  features -  low energy cut-offs,
the  Compton hump and  emission and absorption lines - onto the primary X-ray
emission. The X-ray spectra and, crucially, their variability observed in few
nearby AGN showed that this matter is highly structured with a range of
ionisation states, densities, geometries and locations (\citealt{Turner09},
\citealt{Risaliti2010a}). In this respect, the significant variability of the
absorbing column density (\nhsym) detected in the so called ``changing look''
AGN, i.e. AGN that have been observed   both  in Compton-thin  (\nhsym =$10^{23}
\rm cm^{-2}$) and reflection dominated states (\nhsym  $>10^{24} \rm cm^{-2}$)
(\citealt{Risaliti2002}),  implies that the absorbing material has to be clumpy and at much smaller distance than the conventional obscuring ``torus" with velocity, distance  and size  from the central X-ray source of the same order of those of the Broad Line Region (BLR) clouds.

Up to now, we can count only a few ``changing look'' AGN  where  such  a variability  
has been discovered on time-scales from a few days down to a few hours:
  NGC~1365 (\citealt{Risaliti2005,Risaliti2007,Risaliti2009}), NGC~4388 (\citealt{Elvis04}), NGC~7674 (\citealt{Bianchi05}),
NGC~4151 (\citealt{Puccetti07}), NGC~7582 (\citealt{Bianchi09}),  UGC~4203 (\citealt{Risaliti2010}), NGC4051 (\citealt{Uttley04,Lobban11}) and 1H 0419-577 (\citealt{Pounds04}).  Among them we recall NGC2992 (\citealt{Weaver96}), however for this source one year monitoring with  RXTE (\citealt{Murphy07}) unveiled the presence of short-term flaring activity rather than a change in the covering of the absorber.\\

Within a  project investigating   the occurrence of AGN in a sample of  interacting galaxies, we came across an
interacting system, NGC454,  which was recently observed in the X-ray energy band   with \suzaku, and
$\sim$6 months later with \xmm, and whose main X-ray spectral components present  interesting    variability properties. 
 
 Here we   compare and discuss  the  X-ray observations from \suzaku , \xmm\ and \swift\ that   unveiled that NGC454  can be placed among
those AGN whose absorbing   \nhsym\ is strongly  variable (section \ref{suz_xmm}). The paper is structured as
follows. The interacting system NGC454 is described in \S \ref{ngc454_description}. The X-ray  observations  and
data reduction are summarized in \S \ref{data_red}. In \S \ref{spectral_analysis} we present the   spectral
analysis of both   datasets and the comparison between the   observations, aimed to assess the  nature of the
X-ray absorber.    
Summary and conclusions follow in \S \ref{conclusion}. \\
Throughout this paper, a concordance cosmology with H$_0=71$ km s$^{-1}$
Mpc$^{-1}$, $\Omega_{\Lambda}$=0.73, and $\Omega_m$=0.27 \citep{Spergel2003} is adopted.

\section{NGC454}
\label{ngc454_description}
 
Optical studies  (\citealt{Arp87,Johansson88,Stiavelli98}) of the interacting 
system NGC454  (see Figure \ref{fig:imaging}, right panel)  describe it as a pair of emission
line galaxies consisting of a red elliptical galaxy (eastern component,
hereafter NGC454E) and a blue irregular galaxy (western component, hereafter
NGC454W), at redshift z=0.0122.  The distorted morphology of both these
galaxies, together with the spectroscopic and photometric evidence of a young
stellar population, is a clear sign of the interacting nature of this system. 
Furthermore, three very blue knots (discussed in section \ref{other_src}),
probably Strongren spheres surrounding clusters  of very hot newly formed stars,
are located   (and likely related) to  the south of   NGC454W. {\emph{HST} 
observations of the system, performed with the Wide Field Planetary Camera 2, 
confirmed that NGC454  is in the early stages of interaction
\citep{Stiavelli98}.
The above authors stated also that an important fraction of gas has drifted to
the center of the eastern component, but it has yet not produced any significant
visible star formation activity; a population of young star clusters has
formed around the  western component. \\ The optical spectrum of NGC454E is
consistent with that of a Seyfert 2 galaxy  (although none of the high
excitation lines, e.g. HeII lines, can be seen) while no optical evidence of an
AGN is present in the  spectrum of NGC454W which is  fully consistent with that
of a star-forming galaxy \citep{Johansson88}.

\section{Observations and data Reduction}
\label{data_red}
 
\subsection{\suzaku\ data}
NGC454  was observed on April 29,  2009 by the Japanese X-ray satellite \suzaku\ (\citealp{Mitsuda07}) for a total exposure time of about 130 ksec.
\suzaku\ carries on board four  X-ray Imaging Spectrometers (XIS, \citealp{Koyama07}), with X-ray CCDs at their focal plane, and a non-imaging hard X-ray detector (HXD-PIN, \citealp{Takahashi07}). At the time of this observation only three of the XIS were working: one  back-illuminated (BI) CCD (XIS1) and two front-illuminated (FI) CCDs (XIS0 and XIS3).
All together the XIS and the HXD-PIN cover the
0.5--10 keV and 12--70 keV bands respectively. The spatial resolution of the XIS is $\sim$ 2 arcmin (HEW), while the 
field of view (FOV) of the HXD-PIN is   34  arcmin radius.
Data from the XIS and HXD-PIN  were  processed using v2.1.6.14
of the \suzaku\ pipeline  and applying the standard screening parameters\footnote{The screening
filters all  events  within the South Atlantic Anomaly (SAA)  as well as  with an
Earth elevation angle (ELV) $ < 5^{\circ }$ and  Earth day-time
elevation angles (DYE\_ELV) less than $ 20 ^{\circ }$. Furthermore
also data within  256s of the SAA were excluded from the XIS and within 500s of the
SAA for the HXD. Cut-off rigidity (COR) criteria of $ > 8 \,\mathrm{GV}$ for the
HXD data and $ > 6 \,\mathrm{GV}$ for the XIS were used.}.

\subsubsection{The \suzaku\ XIS analysis}

The XIS data were selected in $3 \times 3$ and $5\times 5$ editmodes using only
good events with grades 0,2,3,4,6 and filtering the  hot and flickering pixels with
the script \textit{sisclean}; the net exposure times are  103 ksec   for each of
the XIS. The XIS  source spectra  were extracted from a circular region of 2.2$'$
radius  centered on the source,  and the  background spectra  were extracted from two
circular regions with the same radius of the source region,  offset from the source and the calibration
sources.   The XIS response (rmfs) and ancillary response (arfs) files were
produced,   using the latest calibration files available, with the \textit{ftools}
tasks \textit{xisrmfgen} and \textit{xissimarfgen} respectively.  The spectra from
the  two FI  CDDs (XIS 0 and XIS 3) were combined to create  a single source
spectrum (hereafter XIS--FI),  while the  BI (the XIS1) spectrum  was kept separate
and fitted simultaneously.    The net 0.5--10 keV  count rates  are: $(0.0117\pm
0.0005)$ cts/s, $(0.0142\pm 0.0005)$ cts/s, $(0.0132\pm 0.0006)$ cts/s for the  XIS0,
XIS3 and XIS1  respectively. We considered data in the range 0.5--10 keV for the XIS--FI    and in the range 0.6--7 keV  for the XIS--BI (because the XIS--BI is optimized for  observing below $\sim 7$ keV). For both the XIS-FI and XIS-BI we ignored the band 1.6--1.9 keV, due to the presence of instrumental calibration
uncertainties.
The net XIS source spectra were then  binned   to a  minimum 
of 50 counts per bin.   

\begin{figure*}
\begin{center}
\resizebox{0.9\textwidth}{!}{
\rotatebox{0}{
\includegraphics{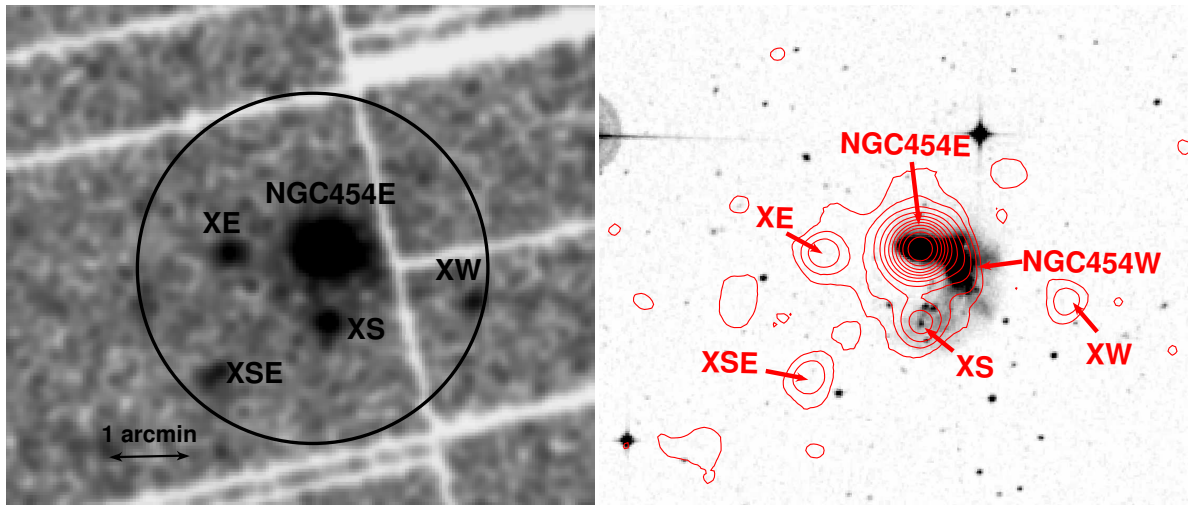}
}}
\caption
{Left panel: \xmm\ EPIC-pn image (0.5--10 keV) with superimposed the \suzaku\ XIS extraction region.     
Right Panel: \emph{Digital Sky Survey} (DSS) optical image with overlaid  the \xmm\ PN 0.5--10 keV contours. We 
 marked the main X-ray sources discussed in the text. It is  evident that the main X-ray source is positionally
coincident with NGC454E (classified as a Seyfert 2) while no X-ray emission is detected at
the position of NGC454W.}
\label{fig:imaging}
\end{center}
\end{figure*}
 
\subsubsection{The \suzaku\ HXD-PIN analysis}
For the HXD-PIN data  reduction and analysis we followed the latest \suzaku\ data
reduction guide (the ABC guide Version
2\footnote{http://heasarc.gsfc.nasa.gov/docs/suzaku/analysis/abc/}),   and used the
rev2 data, which include all 4 cluster units.   The HXD-PIN instrument team
provides the background (known as  the ``tuned'' background) event file, which
accounts for  the instrumental  ``Non X-ray Background'' (NXB; \citealt{kokubun07}).  The
systematic uncertainty of   this ``tuned'' background model is  
$\pm$1.3\% (at the 1$\sigma$ level for a net 20 ksec
exposure\footnote{ftp://legacy.gsfc.nasa.gov/suzaku/doc/hxd/suzakumemo-2008-03.pdf}).\\   We extracted the
source and background spectra   using the same common good time interval, and 
corrected the source spectrum  for the detector dead time. The net exposure time 
after  the screening  was 106 ksec. We  then simulated a spectrum for the cosmic
X-ray background counts \citep{Boldt,Gruber} and added  it to the  instrumental
one.\\  

NGC454 is   detected at a level of  3.4 \% above
the background and the net count rate in the   15--30 keV band is $0.01\pm
0.002$ cts/s. For the spectral  analysis the source  spectrum  was rebinned 
in order to have a signal-to-noise ratio $\geqslant$3 in each energy bin. 
We fit the \suzaku-HXD spectrum with a single absorbed power-law component with a photon index $\Gamma=1.9$  and derived 
an observed 15--30 keV flux of $\sim$3.4$\times 10^{-12}$\flux.

\subsection{The \swift -BAT observation}
NGC454 was also detected with the BAT detector on
board of \swift\ (\citealt{Gehrels04}). BAT is a coded aperture imaging camera  that operates in the 14--150 keV energy range; it has a large field
of view (1.4 steradian half coded), and a point spread function
(PSF) of 18 arcmin (HEW). 
\swift-BAT is devoted mainly to the monitoring of a
large fraction of the sky for the occurrence of gamma ray bursts
(GRBs); while waiting for new GRBs, it continuously collects spectral
and imaging information in survey mode, covering a fraction 
between 50$\%$ and 80$\%$ of the sky every day. \\
NGC454  (BAT  name: SWIFT J0114.4-5522)   is part of the Palermo \swift-BAT 
54-Month hard X-ray catalogue (\citealt{Cusumano10}) and the \swift-BAT
58-Month Hard X-ray Survey (heasarc.gsfc.nasa.gov/docs/swift/results/bs58mon/). 
This last survey 
detected 1092 sources  in the 14--195 keV band down to a significance level of
4.8$\sigma$, reaching a flux level of $1.1 \times 10^{-11}$\flux over 50\% of the sky (and $1.48\times  10^{-11}$\flux over 90\% of the sky); as part of this new edition of the Swift-BAT
catalogue,  8-channel spectra and monthly-sampled light curves for each object detected in the survey were made available
(\citealt{Baumgartner11}).\\  The 14--195 keV observed flux of NGC454  is	1.90$_{-0.5}^{+0.5} \times 10^{-11}$\flux, in agreement, when  accounting for the different bands,      with the flux quoted in the
\citealt{Cusumano10} catalogue ($\rm F_{ 14-195 \ keV}\sim 1.7 \times 10^{-11}$\flux). This flux is also in good agreement
with the expected   14--195 keV flux ($\sim 1.6 \times 10^{-11}$\flux) extrapolated from that measured
with \suzaku\  in the 15--30 keV range .

\subsection{The \xmm\ observation}
NGC454 was observed with \xmm\ on November 5, 2009
 for a total exposure time of about 30 ksec. The \xmm\ Observatory (Jansen et al. 2001) carries, among its onboard
 instruments, three 1500 cm$^2$ X-ray telescopes, each with EPIC (European Photon Imaging Camera) imaging
spectrometers at the focus. 
Two of the EPIC use MOS CCDs
(\citealt{Turner01})  and one uses a pn CCD (\citealt{Struder01}). These CCDs allow observations in the range
$\sim$0.5--10 keV. The spatial resolution of the 2 MOSs is $\sim 14''$ (HEW), and $\sim 15''$ (HEW) for the pn
(\citealt{Ehle01}).

 During this observation the pn, MOS1, and MOS2 cameras had
the medium filter applied and they were operating in full frame Window mode.
The  data have been processed and cleaned using the Science Analysis Software
(SAS ver. 6.5) and analysed using standard software packages (FTOOLS ver. 6.1
and XSPEC ver. 11.3). Event files have been filtered for high-background time
intervals, and only events corresponding to patterns 0--12 (MOS1, MOS2) and
to patterns 0--4 (pn) have been used. The net exposure
times at the source position after data cleaning are $\sim$23.9 ksec (pn),
$\sim$29.1 ksec (MOS1) and $\sim$29.2 ksec (MOS2).

In the right panel of Figure \ref{fig:imaging} we report the optical DSS
image of the system NGC454, together with the  \xmm\  0.5--10 keV contours
(green, in the electronic version only) from EPIC-pn. It is evident that the bulk of the X-ray emission is
positionally  coincident with NGC454E (the galaxy spectroscopically classified  as
a Seyfert 2) while no strong X-ray emission is detected at the position of NGC454W
(the source spectroscopically classified as a star-forming galaxy). We also
detected a weak X-ray source  to the south of NGC454, which is positionally 
coincident with one of the three very blue knots discussed above, likely a star
forming region belonging to NGC454W.

The  pn, MOS1 and MOS2 source spectra  were extracted from a circular region of 0.46 arcmin
radius  centered on the source (NGC454E),  while the  background spectra  were extracted from two
circular regions with 0.5 arcmin radius  offset from the source.   The  MOS1 and MOS2 spectra were combined, then both
the EPIC-pn and  EPIC-MOS  spectra were   grouped with a minimum of 30 counts per channel.

\subsubsection{Contamination from unresolved sources in the \suzaku\ (XIS, HXD) and \swift-BAT extraction region/field of view}
\label{other_src}
In the left panel of Figure \ref{fig:imaging} we show the \xmm \ 0.5--10 keV pn
image  along with the \suzaku \ extraction region (circle with 2.2$'$ radius).
As discussed above the main X-ray source is centered on NGC454E but given the
\xmm\ better angular resolution (14$''$--15$''$ HEW)  as compared to \suzaku\
($120''$ HEW), we can clearly distinguish  4  other X-ray sources, besides NGC454E,
entering in the \suzaku\ XIS extraction region. We extracted the \xmm \ spectra
for the 3 brighter sources (XS, XE and XSE, marked in  Figure \ref{fig:imaging} for clarity)  and
analysed them in order to estimate their possible contribution to the \suzaku\
spectrum; the  remaining source (XW) has only $\sim 80$ counts detected in the $\sim0.5-10$ keV
band (see below).\\

XS is well fitted with a power law, modified only by Galactic absorption, with a
photon index $\Gamma\sim 1.8$  and  a 2--10 keV flux
F$_{\rm[2-10]keV}\sim 1.9 \times 10^{-14}$\flux; the
extrapolated flux in the  14--70 keV band  (assuming $\Gamma\sim 1.8$) is
F$_{\rm[14-70]keV} \simlt3\times 10^{-14}$\flux. As said
above this source is likely associated with a star forming region related to
NGC454W; if so, assuming $z=0.0122$,  its 2--10 keV luminosity is $L_{\rm[2-10]keV} \sim 6.2 \times 10^{39}$ \lum.   We cannot establish if this luminosity is due to one or more sources and thus speculate on its/their nature, because we lack both the spatial resolution and a good enough sampling to assess its variability  and spectral properties.
The source to the east of NGC454E (hereafter XE) can be fitted with a power-law and a thermal component, yielding
$\Gamma\sim$1.6,  $kT\sim 0.3$ and  $F_{\rm[2-10]keV}\sim 8.8 \times 10^{-15}$\flux (F$_{\rm[14-70]keV} 
\simlt 5 \times 10^{-14}$\flux).   The source to the south-east of NGC454E  (hereafter XSE) can be
fitted with an absorbed power law (\nhsym $\sim2.2 \times 10^{22} $ \nh) with photon index set to 1.8 and
F$_{\rm[2-10]keV}\sim 2.7 \times 10^{-14}$\flux (F$_{\rm[14-70]keV}   \simlt 3 \times 10^{-14}$\flux). 
Finally, the fourth source located to the west of 
 NGC454E (hereafter XW) has not enough counts for a meaningful spectral analysis ($\sim 80$  counts) and  its estimated
 fluxes are F$_{\rm[2-10]keV}\sim$2.0$\times 10^{-14}$\flux  and 
 F$_{\rm[14-70]keV}  \simlt 2.5\times 10^{-14}$\flux  (adopting $\Gamma\sim 1.9$)}. According to the extragalactic logN-logS distributions computed by \cite{Mateos08}, at this flux level the number of random  2--10 keV sources  in the \suzaku\ extraction region is $\sim 2$, thus the sources XE, XSE and XW 
are probably those expected by "chance". There is no NED identification available  for XE, XSE and XW.\\

The combined 2--10 keV flux   of all these 4 possible contaminating sources
(F$_{\rm[2-10]keV}\sim 7.5\times 10^{-14}$\flux) imply that they will provide a
negligible contribution to the \suzaku\ XIS spectrum of NGC454
(F$_{\rm[2-10]keV}\sim 6\times 10^{-13}$\flux).  More important their estimated
F$_{\rm[14-70]keV}$ are well below the \suzaku\ HXD-PIN   or   Swift-BAT
sensitivity. On the other hand this check is still not sufficient for these two
latter instruments since their FOV is  larger than that of the \suzaku\ XIS
instrument. Assuming that the X-ray emission above
10 keV detected with    the HXD-PIN or the Swift-BAT is associated to  the same
source, as the good agreement  of the measured fluxes strongly suggests, we can use
the instrument with the smaller FOV (\swift-BAT) to perform further checks. In
particular using known catalogues or archives  (NED\footnote{http://ned.ipac.caltech.edu/} and SIMBAD\footnote{http://simbad.u-strasbg.fr/simbad/}) we searched for
bright X-ray/optical sources within 6 arcmin  radius error circle (corresponding to 99.7\% confidence level
for a source detection at 4.8 standard deviations, \citealt{Cusumano10}) that could be responsible of the
observed X-ray emission above 10 keV.  No  plausible contaminant source was found and,
in the following, we will assume that the emission above 10 keV comes from
NGC454E. We note that we are also assuming a negligible contribution to the
emission above 10 keV from the companion galaxy in the interacting system, NGC454W.
While a confirmation of this assumption has to wait for direct imaging
observations  above 10 keV with adequate spatial resolution,   we stress that no
emission  was detected below 10 keV from NGC454W, while a contribution would be
expected even in the  case of a deeply buried AGN (see e.g. \citealt{Dellaceca2002}).
We thus conclude that    we do not expect significant
contaminations from the nearby sources to the \suzaku\ and \swift\ spectra.

\section{Spectral analysis}
\label{spectral_analysis}

\subsection{The \suzaku\   and \swift\ broad band   X-ray emission}
\label{suzaku_spectra}

We  first considered the X-ray spectrum of NGC454E in the 0.5--100 keV band by
fitting simultaneously the  \suzaku\ XIS, \suzaku\ HXD and \swift-BAT data. The
cross-normalisation factor between the HXD and the XIS-FI was set to 1.18, as
recommended for   HXD nominal observation processed after 2008 July (Manabu et al.
2007; Maeda et al.
2008\footnote{http://www.astro.isas.jaxa.jp/suzaku/doc/suzakumemo/suzakumemo-2007-11.pdf;\\
http://www.astro.isas.jaxa.jp/suzaku/doc/suzakumemo/suzakumemo-2008-06.pdf}),
while the cross-normalisation between \swift\ and XIS was allowed to vary. \\ In
the   subsequent sections  the $\chi^2$ statistics was used for the fit, the 
errors are     quoted to 90\% confidence level for 1 parameter of interest and
all the    spectral parameters  are quoted in the rest frame of the source. \\

We    fitted the continuum with a redshifted unabsorbed power-law model,
modified only by Galactic  (\nhsym $=2.73 \times 10^{20}$ \nh,
\citealp{Dickey1990})  absorption.  This model did not provide an adequate
description of the broadband spectrum of NGC454E ($\chi^2$/dof=522.6/122). If we
fit only the 2--5 keV continuum, excluding   possible complexity in the soft
energy range and  near   the   Fe K  emission line complex   we found a
very flat photon index ($\Gamma \sim 0.15$),  strongly suggesting that we are
dealing with an absorbed AGN, in agreement with the optical
spectral classification of NGC454E.  \\
\begin{figure}
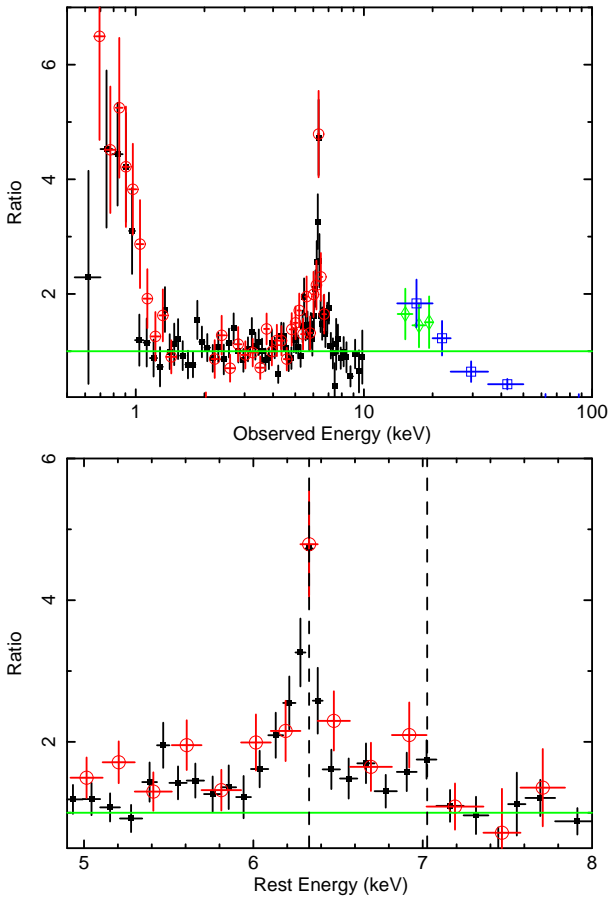

\begin{center}
\resizebox{0.46\textwidth}{!}{
\rotatebox{-90}{
\includegraphics{marchese_fig2.ps}
\includegraphics{marchese_fig3.ps}
}
}
\caption{Upper panel: ratio between the \suzaku\ and \swift\ data (XIS-FI: black filled squares; XIS1: red  open circles; HXD: green rhombs; and BAT:blue open squares, colors in the electronic version only)
and the unabsorbed power-law model used to fit \suzaku\ data in the 2--5 keV energy
range. Lower panel: zoom into the  5--8 keV energy range (XIS-FI: black filled squares, XIS1:  red open  circles).   We can clearly see  at 6.4 keV the excess characteristic of the
Fe K$\alpha$  emission line  and at $\sim 7$ keV, the combined contribution of the Fe K$\beta$  emission
line (7.06 keV), Fe XXVI ($\sim$6.97 keV), and the reflector edge. The central energies of the Fe K$\alpha$ and Fe K$\beta$  are marked with dashed vertical lines.}
 \label{fig:res1}
\end{center}
\end{figure}

\begin{figure}
\begin{center}
\resizebox{0.46\textwidth}{!}{
\rotatebox{-90}{
\includegraphics{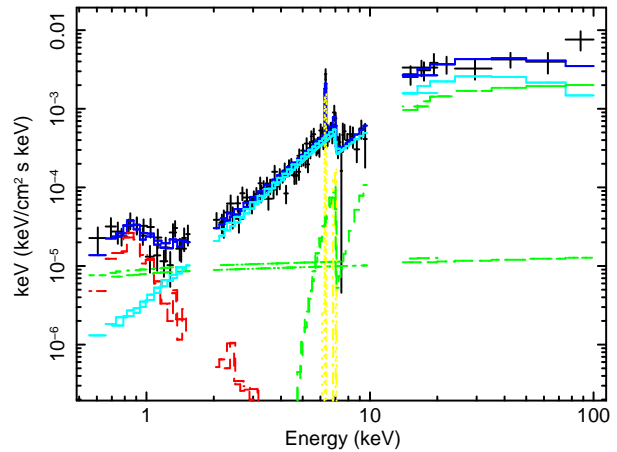}
}}
\caption{ Unfolded \suzaku\ spectrum, showing separately the different components of the best-fit model: in green (in the electronic version) the scattered power-law and the primary absorbed power-law,  in red  (in the electronic version) the soft thermal component, in yellow  (in the electronic version) the iron K$\alpha$ and K$\beta$ emission lines, in light blue  (in the electronic version) the reflection component, and in blue  (in the electronic version) the total resulting spectrum.}
\label{unf_spectrum}
\end{center}
\end{figure}

The  residuals with respect to
this simple unabsorbed power-law model, which   are shown in Figure
\ref{fig:res1},  allowed us  to infer the main
features of the observed spectrum. An excess at energies below 1 keV, an emission
line feature at $\sim$6.4 keV (likely associated with Fe K$\alpha$),
together with a  line-like feature at $\sim$7 keV, and an excess at energies
between 10 and 20 keV are clearly evident. The residuals in the soft X-rays
suggest the presence of a  thermal component probably related to the host galaxy. The
simultaneous occurrence of a strong Fe K$\alpha$ emission line  at  $\sim$6.4 keV
(figure \ref{fig:res1} upper and lower panel), a very flat observed $\Gamma$  and 
an excess in the hard X-rays (above 10 keV) is the distinctive spectral signature
of  a highly absorbed source, with a possible  strong Compton
reflected component. The excess observed at $\sim$7 keV (Figure
\ref{fig:res1} lower panel) is likely due to the  combination of    the Fe K$\beta$
emission line (7.06 keV) and the  Fe XXVI ($\sim$6.97 keV) and the Fe edge ($\sim
$7.11 keV).\\

Given these features, we   performed a broadband fit including: 
\begin{enumerate}
\item a thermal component (modelled with the \textsc{mekal} model, \citealp{mekal});
\item an absorbed primary power-law component;
\item an unabsorbed power-law component with the same photon index $\Gamma$;
\item  two  Gaussian emission lines at $\sim$6.4 keV (Fe K$\alpha$) and 7.06
keV (Fe K$\beta$) respectively.  We kept the energy of the  Fe K$\beta$ fixed to
7.06 keV, tied its intrinsic width ($\sigma$) to   the  width of the
corresponding Fe K$\alpha$ line and fixed its flux to be  13\%   of the Fe K$\alpha$ flux,
consistent with the theoretical value (\citealp{Kaastra93});
\item a Compton reflected component, modelled with the \textsc{pexrav} model in 
\textsc{xspec} (\citealp{pexrav}). The parameters of the reflected component are:
an inclination angle i fixed to 63$^{\circ}$, abundance Z=Z$_{\odot}$,  a reflection
fraction (defined by the subtending solid angle of the reflector $R=\Omega/4\pi$) 
fixed to be -1 (i.e. pure reflection)\footnote{Since in the ``pure reflection'' \textsc{PEXRAV} model there is a
degeneracy between R and the normalisation, we set the reflection scaling factor
to -1   and allowed the normalisation to vary.},  the cut-off energy (fixed at 200 keV, \citealt{Dadina2008}) and the normalisation.
\end{enumerate}
 The absorber  was
 modelled by a combination of the \textsc{cabs} and \textsc{zphabs} models
in \textsc{XSPEC}, assuming the same column density, since they represent the same medium producing two different effects (i.e. the  non-relativistic 
Compton scattering and photoelectric absorption of the primary radiation, respectively). 

The model setup is:
\begin{center}
\textsc{wabs$\times$[ mekal  + zpowerlw  + zgauss  + zgauss  + pexrav  + cabs$\times$zphabs$\times$( zpowerlw)]}
\end{center}

We found that this model provides a good representation of the  X-ray
emission of NGC454E ($\chi^2$/dof=104.5/103).  The resulting best-fit
parameters are reported in table \ref{tab1}.
In particular, this best-fit model yielded $\Gamma=1.92_{-0.36}^{+0.29} $,
\nhsym $=2.05^{+4.25}_{-1.38}\times 10^{24 }$ \nh. The rest-frame energy of the Fe
K$\alpha$ is $E_{K\alpha}=6.38 \pm 0.02$  keV and its equivalent width   with
respect to the observed continuum is EW=340$^{+60}_{-80}$eV. At the \suzaku\ spectral
resolution this emission line is  unresolved;  leaving  the          width
$\sigma$ free to vary we found  $\sigma \lesssim$ 70 eV (at the 90\%  confidence
level), thus we fixed it to be $ \simeq 10$ eV.  The cross-normalisation factor
between  the \swift-BAT  and  the XIS-FI is 1.05$^{+0.64}_{-0.39}$.   We stress that a different choice of the cut-off energy in the range between 100 and 300 keV does not affect significantly the best-fit reflection parameters obtained in this work.
f The relative importance of the reflection component is given by the ratio  
between the normalizations of the primary absorbed power-law and the reflection component; in
our case  this ratio is $\sim$0.5, which at first order would correspond to a  reprocessor covering  a solid angle  2$\pi$. The   fraction of scattered radiation is $\sim 1\%$.  The observed 2--10 keV flux is $\sim 6.3
\times 10^{-13}$\flux while the   intrinsic 2--10 keV luminosity obtained with this
model   is 7.2$\times 10^{42}$\lum.



\begin{table*} 
\caption{Summary of the \suzaku\   and \xmm\  parameters for the best-fit  models described in section
\ref{suzaku_spectra}, and \ref{ion_abs}. 
\label{tab1}
}
 \begin{tabular}{lccc }
\hline
 Model Component  &  Parameter  &  \suzaku &\xmm\ \\ 
 \hline
&&   \\
Power law &$\Gamma$&$1.92_{-0.36}^{+0.29}$ & $1.99_{-0.07}^{+0.11}$ \\
& Normalisation$^{a}$ & $ 7.39_{-4.39}^{+30.00,b}   $ & $ 2.77_{-0.65}^{+0.71} $\\

Scattered Component  &Normalisation$^{a}$ &$8.55_{-4.52}^{+5.48}\times 10^{-3}$ & $1.62_{-0.22}^{+0.29}\times 10^{-2}  $\\
Absorber   & N$_{\rm H}$& $2.05_{-1.38}^{+4.25}\times 10^{24}$ \nh & $1.0^{+0.1}_{-0.2}\times 10^{23}$ \nh \\
Thermal emission  & kT& $0.62_{-0.17}^{+0.10}$ keV &$0.62_{-0.11}^{+0.11}$ keV \\
	  &Normalisation$^c$ &$ 6.94^{+2.40}_{-2.22} \times 10^{-6}  $   & $3.49 ^{+1.52 }_{-1.50 }  \times 10^{-6}   $   \\

Neutral reflection & Normalisation$^a$ &$ 3.46^{+2.14}_{-1.61 } $ &  $3.55^{+1.52 }_{-1.81} $  \\

Fe K$\alpha^{d}$ &   Energy    & $6.38_{-0.02}^{+0.02}$ keV &$6.36 _{0.03} ^{+0.03}$ keV    \\
          &   EW        & 340$^{+60}_{-80}$ eV & $120^{+40}_{-40}$ eV\\
          & Normalisation$^e$ & 3.62$^{+0.79}_{-0.78} \times 10^{-3}$ &  4.75$^{+1.35}_{-1.40} \times 10^{-3}$   \\
Ionised Absorber &  N$_{\rm H}$ & .. & $6.05^{+8.95}_{-4.10} \times 10^{23}$\nh \\   
               & log$\xi$ & ..&$3.55^{+0.49 }_{-0.25} \rm erg \ cm \ s^{-1}$\\      
               	& $v_{turb}$& ..& 300 km $\rm s^{-1}$   \\
   
    & $\chi^2$/dof   &   104.5/103 &190.7/197  \\
&F $_{(0.5-2)\mathrm {keV}}$ &$\sim 4.9\times 10^{-14}$\flux &$\sim  5.8 \times 10^{-14 }$\flux\\
&F$_{(2-10)\mathrm {keV}}$  &$\sim 6.3 \times 10^{-13}$\flux &$\sim 1.9  \times 10^{-12 }$\flux\\
&F $_{(14-150)\mathrm {keV}}$ &$\sim 1.4 \times 10^{-11}$\flux &$\sim 1.3   \times 10^{-11 }$\flux\\
&L $_{(0.5-2)\mathrm {keV}}$  &$\sim 4.7 \times 10^{42}$\lum&$\sim 2   \times 10^{ 42 }$\lum \\
&L$_{(2-10)\mathrm {keV}}$  &$\sim 7.2\times 10^{42}$\lum &$\sim 2.5   \times 10^{42}$\lum\\
&L$_{(14-150)\mathrm {keV}}$  &$\sim 1.4 \times 10^{42}$\lum &$\sim 4.8  \times 10^{ 42}$\lum \\
 \hline
\end{tabular}\\
\begin{flushleft}
 
$^a$ units of $10^{-3}$ photons  keV$^{-1}$ cm$^{-2}$  s$^{-1}$.\\
$^b$   Due to a degeneracy between the normalisations of the primary power law  and   \textsc{pexrav}, the errors were computed fixing the reflection normalisation to its best-fit value.\\
$^c$  The  normalisation of the thermal component is defined as $K=\frac{10^{−14}}{4\pi(D_A (1 + z))^2} \int{n_e n_H dV}$ 
where $D_A$ is the angular diameter distance, z is the redshift,\\ $n_e$ and $n_H$ are the electron and hydrogen density (cm$^{-3}$) respectively, and dV is the volume from which the deprojected emission originates.  \\ 
$^d$ The line is unresolved; the intrinsic width has been fixed to be  $\simeq$ 10 eV.\\
$^e$ units of  $10^{-3}$ photons   cm$^{-2}$  s$^{-1}$. \\
\end{flushleft}
\end{table*}
 
\subsection{Comparison with \xmm\ data}
\label{suz_xmm}
 
\begin{figure}
\begin{center}
\resizebox{0.46\textwidth}{!}{
\rotatebox{-90}{
\includegraphics{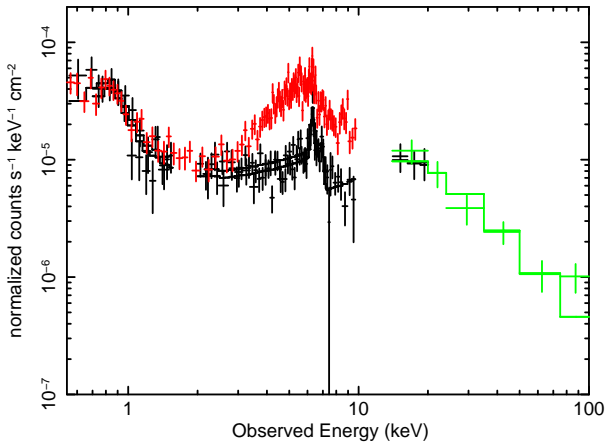}
}}
\caption{Comparison between the \suzaku\ XIS (black, lower spectrum), HXD (black), \swift-BAT (green, colors in the electronic version only) and the
\xmm\ (red, upper spectrum) data showing the dramatic change in the curvature in the 3--6 keV energy range. The underlying model
(black  and green line) is the one obtained fitting only the \suzaku\ XIS (black)  and \swift-BAT data (green). 
}
\label{fig2}
\end{center}
\end{figure}
In Figure \ref{fig2} we report the \suzaku\ XIS (black, lower spectrum), HXD (black) and \swift-BAT
spectra (green, in the electronic version only). In red (upper spectrum) we also show the \xmm\ pn and MOS
data, revealing a dramatic change in the spectral  curvature   between 3 and
6 keV. This variation   is most likely  due to a
change in the  amount of absorption of the primary radiation. 
To test this hypothesis we applied the \suzaku\ best-fit model to the \xmm\
spectra, leaving     only the absorbing column density (\nhsym)  free to vary. 
We also left  both the  cross-normalisation factors  between   the  pn  and
the MOS  spectra and  between \swift-BAT and pn data free to vary; they were found to be 1.02$\pm 0.04$   and
1.06$^{+0.14}_{-0.16}$ respectively.
During the \xmm\ observation   the  \nhsym\ decreased by  about one order of
magnitude (from $\sim 2.1 \times 10^{24}$\nh \ to  $\sim 2.6 \times 10^{23}$\nh); 
 this change in the amount of absorption is sufficient to explain  the bulk of the differences between the observed \xmm\ and \suzaku\ spectra.
 \\ For completeness, we also allowed to  vary the photon index
($\Gamma$), the normalisation of  both the  power law components, the thermal
component  and the K$\alpha$ energy and normalisation. The fit yielded
$\chi^2$/dof=213.9/199 and the  only parameter changing well beyond the \suzaku\
errors is, as expected, the \nhsym, decreasing to $2.78^{+0.16}_{-0.17}  \times
10^{23} $ \nh. 
  This confirms that the strong variation between \xmm\ and \suzaku\
is due to a change in column density of $\Delta$\nhsym$\sim$1.8$\times 10^{24}$ \nh .  \\
  Prompted from this result we also inspected    the 3 \swift-X-ray Telescope (XRT) observations taken in 2006  with a time lag of the order of 1--2 days from each other, and we found that the source was in a state similar to that observed by \xmm. The exposure time of each of the observations is  less than 10 ksec (8713, 8661 and 3667  sec respectively), thus the relatively low  statistics does not allow us to establish if there is a variability  between  the single observations.

\begin{figure}
\begin{center}
\resizebox{0.46\textwidth}{!}{
\rotatebox{-90}{
\includegraphics{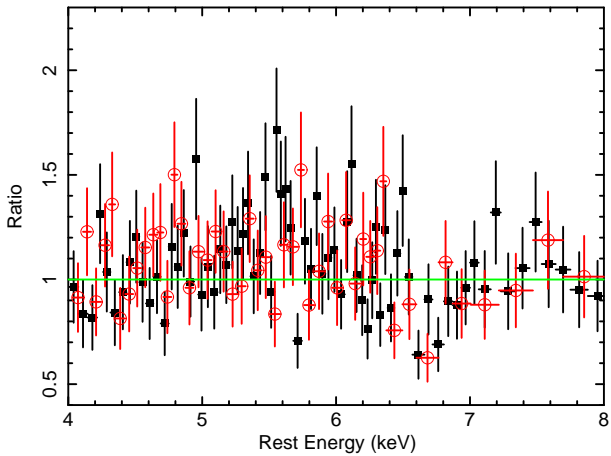}
}}
\caption{Residuals of the \xmm\ data (pn data are the black filled squares  and MOS data are the red open circles) in the range 4--8 keV with respect to the spectral model discussed in section
\ref{suz_xmm}. An absorption feature at about 6.7 keV and a 
spectral curvature in the 5--6 keV range are clearly present.}
\label{fig:res2}
\end{center}
\end{figure} 

A closer inspection of the residuals in the 5--8 keV energy range to this  best-fit model showed   some residual
curvature between 5 and 6 keV, together with a possible absorption feature centered at $\sim $ 6.7 keV 
(see Figure \ref{fig:res2}),  which is present in both the  pn and MOS spectra and
is suggestive of a more complex and likely ionized absorber. 
After checking the significance of this absorption line, we included in the model an
additional ionized absorber (see \S \ref{ion_abs}).   We note that, after accounting for the
absorption feature at $\sim$6.7 keV, the excess of curvature in the range 5--6 keV
is not present anymore. \\

The parameters of the \xmm\ best-fit  model are reported in table
\ref{tab1} and the final model setup is described in section 
\ref{ion_abs}. We note that the difference in the best-fit normalisations
of the thermal and scattering component between \suzaku\ and \xmm\ are likely due to
a degeneracy between these two parameters. Indeed  the 0.5--2  keV  flux  did not
strongly vary between the two observations.\\   The  fraction of scattered radiation is $\sim 5\%$.  The
2--10 keV flux is $\sim 1.9  \times 10^{-12 }$\flux, while the 2--10 keV
luminosity, $L_{(2-10)\mathrm {keV}}$, is $\sim 2.5  \times 10^{42}$\lum, about a
factor 2.8  below the luminosity computed using only the \suzaku\ data.  Although such a variation of the intrinsic luminosity is not unusual in AGN, part of   this  difference could be due to the geometry assumed by the models adopted for the high column density absorber.
Indeed we will show in  section \ref{mytorusMODEL} that this difference
is smaller (a factor 1.7) when we adopt the {\it{Mytorus}} code for the absorber.

\subsubsection{The $\sim 6.7 $keV absorption feature   in the \xmm\   observation}
\label{ion_abs}

 As a first step to model the absorption feature in the 6--7 keV band we added a Gaussian  absorbing component; setting $\sigma =0.05$ keV we
found that the centroid of the line is at E=$6.75^{+0.06}_{-0.04}$ keV, the normalisation is
$-3.75^{+1.12}_{-1.13}  \times 10^{-6}$  and $\Delta \chi^2$=24 for 2 dof. The absorption lines appears to be
marginally resolved; however  leaving its width free to vary we can set only an upper limit $\sigma<0.3$ keV, 
while the   energy centroid is found to be consistent within the errors
E=$6.77^{+0.08}_{-0.06}$keV. The energy of this absorption  line suggests an
association with  absorption  from highly ionized Fe   (i.e.  \fexxv\  at E$\sim$6.7
keV) and thus a clear signature of the presence of an ionized absorber. The presence
of  an ionized absorber is not  exceptional since recent sensitive
observations with \chandra, \xmm, and \suzaku\  unveiled the presence of  red- and blue-shifted
photoionized absorption lines both in  type 1 and type 2 AGN as well as in Radio Quiet and Radio Loud AGN 
\citep{Tombesi2010b,Tombesi2011}.
 Thus,  it appears that
there is a   substantial amount of  ionized gas in the nuclei of AGNs, which may be
linked  to  gas outflowing on parsec scales with velocities from hundreds of km/s 
up to  $v_{\rm out}\sim 0.04 -0.15c$ (\citealt{Tombesi2010b}). We note that
   red- and blue-shifted absorption lines are predicted in several
theoretical models of failed disk winds \citep{Proga2004,Sim2010} or of aborted jet 
\citep{Ghisellini2004}. However, before proceeding with any further modeling of the  absorption feature we
checked its significance. \\





To assess the significance of the  absorption feature  we performed extensive Montecarlo simulations as  detailed
below.  We assumed as our null hypothesis model the best-fit  model discussed at the end of section \ref{suz_xmm},
and we simulated  S=3000  spectra (with the \emph{fakeit} command in \textsc{XSPEC}), with the same exposure
time as the    real data. Each one of these simulated spectra was then fitted with the null hypothesis model to
obtain a $\chi^{2}$ value, and we systematically searched for an absorption line  in the
2--10 keV energy range, stepping the energy centroid of the Gaussian in increments of 0.1
keV and refitting at each step. We then obtained for each simulated spectrum a minimum $\chi^{2}$ and created a
distribution 3000 simulated values of the $\Delta \chi^2$ (compared to the null hypothesis model). This
indicates the fraction of random generated absorption features in the 2--10 keV  band that are
expected to have a $\Delta \chi^2$  greater than a threshold value. If N of these simulated values are greater
than the real value, then the estimated detection confidence level is 1-N/S. Using this analysis we can then
conclude that the line detection is significant at $>$99.97$\%$ level.\\

In order to obtain a
  physical description of the absorber we replaced the Gaussian absorption line with a model representing  a 
photoionized absorber, which  has been produced using a   multiplicative grid of absorption model    generated with the 
{\sc xstar}v 2.1 code (\citealp{xstar}). This grid describes an ionized absorber
parametrised by its column density (\nhsym), and its ionisation parameter, defined as: 
\begin{equation}
\label{ion_abs_eq}
\xi=\frac{L_{ion}}{nR^2}
 \end{equation} 
 where $L_{ion}$ is the ionising luminosity between 1--1000 Rydbergs (13.6
eV to 13.6 keV), \emph{n} is the hydrogen gas density in cm$^{-3}$ and $R$ is the radial distance of the absorber
from the ionising source.   Since there is no apparent broadening of the absorption line we assumed a   
turbulence velocity    of $v_{\rm turb}=300 \ \rm km \ s^{-1}$.  \\ {The inclusion
of this ionized absorber significantly improved  the fit ($\chi^2$/dof=190.7/197,
$\Delta \chi^2 =25$ for 2 dof), with a column  density of \nhsym\ $=
6.05^{+8.95}_{-4.10} \times 10^{23}$ \nh and an ionisation of log($\xi/\rm erg \ cm
\ s^{-1})=3.55^{+0.49 }_{-0.25}$.   The improvement in the $\chi^2$ is determined solely by  fitting the absorption feature in the 6--7 keV band, since an ionised absorber with such a high level of ionization does not produce any feature in the soft band of the continuum.\\

  The parameters of the \xmm\ best-fit  model are reported in table \ref{tab1} and the setup is the following:
\begin{center}
\textsc{ wabs$\times$[ mekal  + zpowerlw  + zgauss  + zgauss  + pexrav  + XSTAR *cabs *zphabs $\times$(zpowerlw)]}
\end{center}

We can now estimate what is the maximum distance of this ionised absorber from the central black hole, using equation \ref{ion_abs_eq}, relating the ionisation parameter, the density of the absorber and the continuum luminosity $L_{ion}$. In this case $L_{ion}$ (in the energy range between 13.6
eV and 13.6 keV) is $7.3 \times 10^{42}$ erg\ s$^{-1}$. Assuming that the thickness of the absorber $\Delta$R=$\rm N_{H}/n$ is  smaller than the distance R$_{ion}$ ($\Delta$R/R$_{ion}$ $<$ 1), we can set an upper limit to the distance:
\begin{equation}
R_{ion}=\frac{L_{ion} \Delta R}{N_H \xi R} <\frac{L_{ion}  }{N_H \xi}=2.3 \times10^{15} \rm cm
\end{equation}
This maximum distance of $\sim 10^{-3}$pc is  consistent with a location of  the 
ionised absorber   within the Broad Line Region of the AGN.   Indeed an estimate of the BLR  size R$_{BLR}$ for NGC454E can be inferred by using the relation   between R$_{BLR}$ and the monochromatic luminosity at 5100 \AA, $L_{5100 \AA}$ (\citealt{Kaspi05}, $ \frac{R_{BLR}}{10 \ lt-days}=2.45 \times (\lambda L_{\lambda}(5100\AA))^{0.608}$).  Since the luminosity of the optical continuum cannot be measured directly from the spectrum, because of the strong absorption, we estimate   $L_{5100 \AA}$ from the intensity of the [OIII]5007\AA \ line flux, assuming a mean F[OIII]5007\AA/F(5100\AA) ratio. 
This ratio has been inferred from the AGN template presented in \cite{Francis91} ($F(5100\AA)=0.059 F([OIII])$). Using the [OIII]5007\AA  \ flux published in Johansson (1988) we obtain  $L_{5100 \AA} \sim 1 \times10^{41}\rm erg \ s^{-1} \ \AA^{-1}$ and, thus, an approximate size of the BLR of 0.05 pc, i.e. about 50 times $R_{ion}$.  \\ 

Since we do not
observe this absorption feature in the \suzaku\ spectrum we added to the \suzaku\
best-fit model a gaussian absorption line with the same parameters obtained with the
\xmm\ data. The lower limit for the detection of an absorption line   with central
energy of 6.75 keV and width of 0.05 keV, is -1.18$\times 10^{-6}$  for \suzaku\
data. Thus, being  the normalisation of this line -3.75$\times 10^{-6}$ in the \xmm\
spectrum, we can infer that the ionised absorber should be detectable by  \suzaku.
The simplest interpretation is that  also the ionized absorber is variable;  
which is   not surprising  since there are several  reported cases of variable
absorption feautures  \citep{Tombesi2010b,Tombesi2011}  (\citealt{Braito2007}; \citealt{Cappi2009}; \citealt{Dadina2005}; \citealt{Risaliti2005}). 
Moreover    instability of the outflowing ionized absorbers  is predicted both
in   disk winds models (\citealt{Proga2004}; \citealt{Sim2010}) or of aborted jet
\citep{Ghisellini2004}. This will cause the presence of  transient absorption
features and variability of the derived outflowing velocities and their EW as observed
in several sources (see e.g. \citealt{Tombesi2010b})


\subsection{A physical interpretation with {\it{Mytorus}} model}
\label{mytorusMODEL}

The models discussed so far, which are based on spectral components largely used
from the
astronomical community, do not treat both fluorescent lines and continuum components
self-consistently. Furthermore all these spectral components may be deficient in one
or more aspect
of modelling the complex transmission and reflected spectrum of AGN over a broad
energy range and for
a large range of absorbing column densities (see section 2 of Murphy and Yaqoob,
2009 for a critical
discussion of these points).

In order to alleviate these problems and, thus, to further assess the  possible
geometry
and/or nature of the variable absorber, we tested the most recent model for the
toroidal
reprocessor  \footnote{http://www.mytorus.com/}  \citep{Mytor}.
This model, recently included in the XSPEC software package, is valid for column
densities
in the range $10^{22}$ to $10^{25}$ cm$^{-2}$ and for energies up to 500 keV (the relativistic
effects
being taken into account); more importantly the reprocessed continuum and fluorescent
line emission
are treated self-consistently for the first time.   This model assumes    that the absorber geometry is toroidal with an opening angle of 60$^\circ$.
Since we are clearly seeing a variation of the absorbing column density along the
line of sight
we have used a spectral configuration of {\it{MyTorus}} that can {\it mimic} a clumpy absorber
and  which also takes into account the fact that the Fe K$\alpha$ is rather
constant (see table \ref{tab1}).\\

We have done this by decoupling the line-of-sight   continuum passing through
the reprocessor   (or  zeroth order continuum, see http://www.mytorus.com/manual/index.html)  and the   reflected  (or  scattered continuum, see mytorus model)  continuum   from reprocessor. 
In practice we allowed the column densities of the line-of-sight continuum and
scattered-reflected continua to be independent of each other.
The  reflected  continuum, and the fluorescent line emission which is
consistently produced in the same location, is not extinguished by another column of
intervening matter. Since the \xmm\ spectrum unveiled the presence of an
additional ionized absorber,  which affects the line-of-sight continuum we also
included  an
ionized absorber, which is modelled adopting the same XSTAR grid as described
in section \ref{ion_abs}.  In  order to do that, we   disentangled the absorbing column density
of the line-of-sight (los) component  from that for the scattered continuum 
plus fluorescent emission lines. The inclination angle of the los component has been
fixed at 90 degrees; the inclination angle for the  reflected/ scattered continuum plus line component component
was, for simplicity, fixed at 0 degrees since the effect of the inclination angle on
the shape of the scattered continuum is not sufficiently large when the scattered
continuum is observed in reflection only.
 Physically, the situation we are modelling by means of this decoupling could
correspond to a patchy reprocessor in which  the scattered continuum is observed
from reflection in matter on the far-side of the X-ray source, without intercepting
any other ``clouds,'' , while the intrinsic continuum is filtered by
clouds "passing" through our line-of-sight to the central engine.

We applied this model   to  the  the  XMM-EPIC  and  \swift-BAT spectra
and we found a god fit with the same   absorbing column density ($N_{\rm {H}}= 2.75^{+0.05}_{-0.04}
\times 10^{23}$\nh; $\chi^2/$dof=$201/192$) filtering the line-of-sight intrinsic
continuum and producing the scattered component (including the production of the fluorescent
emission lines).
The parameters of  the ionized absorber are: $N_{\rm{H}}= 6.46^{+5.04}_{-1.96}\times
10^{23}$\nh and  log$\xi=3.26 ^{+0.20}_{-0.21}$\logxi; these values are in
good agreement with those found with the best-fit model described in table
\ref{tab1}. The photon index of the primary power-law component is now
$\Gamma=1.86^{-0.11}_{+0.17}$  and the intrinsic emitted luminosity
is L$_{[2-10]{\rm keV}}\sim 1.4\times 10^{42}$\lum.
Using the \suzaku\  and \swift\ data, we found that a good fit can be obtained
with an absorber producing the reflected components having an $N_{\rm H}$
statistically consistent with that
obtained using the \xmm\ data (thus suggesting that this component is likely
associated with the distant reflector or torus), while  our line of sight to the
central engine intercepts a  column density  $N_{\rm H}=(0.88\pm0.09)
\times 10^{24}$ \nh.

In summary this analysis,  which is based on a model which takes into account
consistently the
physical process in place within the X-ray absorber, shows that the change of state
of NGC454E can
be understood simply by a chance change in the line-of-sight obscuration
($\Delta N_{\rm H}\sim 6 \times 10^{23}$ \nh) while the global obscurer remains
unchanged. 
The intrinsic luminosity derived from the \suzaku\  data is L $_{[2-10]
\rm{keV}}\sim 2.4\times 10^{42}$\lum . 
We note that using {\it{Mytorus}} the derived  change of the intrinsic luminosity
between the two data sets are in better agreement (a factor 1.7) with respect to
those found in section 4.2.    However, with the present statistic and the complexity of the observed  spectra, we cannot rule out or confirm a possible variation in luminosity of about a factor 2,  frequently observed in AGN; indeed by comparing   the 54-months and 9-months BAT high energy (14--195 keV) spectra of this source, we found that
the intensity is higher in the 9-months  spectrum. However, fitting the spectra with a single absorbed power-law component, we found that a constant flux is also  well within the errors on the best-fit normalizations of this primary power-law.


\section{Summary and Conclusion}
\label{conclusion}

We have presented the results of \suzaku, \xmm\ and \swift\
observations of the interacting system NGC454
(z=0.0122). The
bulk of the measured  2--10 keV   emission comes from the active galaxy NGC454E (L
$_{[2-10]
\rm{keV}}\sim 2\times 10^{42}$\lum); no emission from the center of the companion
galaxy (NGC454W) in the
interacting system is detected. The nuclear X-ray emission of NGC454E is filtered by an absorbing column density typical
of a Seyfert 2 galaxy, in agreement with the optical classification.\\

A comparison between   \suzaku\ and \xmm\ observations (taken 6 months later)
revealed a significant change in the
spectra of NGC454E in the energy range between 3 and 6 keV.   This
variation can be well explained by a variability of
about an order of magnitude in the absorbing column density along the line of sight:
from $\sim 1 \times 10^{24}$\nh
(\suzaku) to $\sim 1\times 10^{23}$\nh (\xmm). This study also adopted  the most recent model for the toroidal reprocessor   (\citealt{Mytor}),  which takes into account consistently the physical processes in place within the X-ray absorber. Furthermore, regarding the \xmm\
spectrum, we detected  a statistically
significant absorption feature a  $\sim$ 6.7 keV, a clear signature of the presence
of a ionised absorber, with
ionisation parameter    log($\xi/\rm erg \ cm
\ s^{-1}) = 3.55$ and column density  \nhsym\ $= 6.05  \times
10^{23}$ \nh. The absence
of this feature in the \suzaku\ spectrum, despite its detectability, implies that it
has varied between the two observations.
Absorption lines associated with ionized iron have been now
observed in several sources and   there is also a  clear evidence
that these lines are variable as in the case of NGC454. Furthermore, in some cases
the measured blue-shifts of the energy centroids imply a large velocity of these
absorbers and a likely association with
powerful disk winds (King \& Pound 2003),  while in other cases
there is no measurable motion as in our case.\\

In summary, with respect to the absorbing column density
variability, NGC454E is a new member of the class of ``changing
look'' AGN, i.e. AGN that have been observed in both Compton-thin
(\nhsym =$10^{23} \rm cm^{-2}$) and reflection dominated states
(\nhsym  $>10^{24} \rm cm^{-2}$). A possible scenario is that a stable and likely distant absorber  responsible
for the iron emission line is present.  However,  there is also a  clear variation
of the \nhsym\  of the line of sight absorber, probably indicative
of the clumpy nature of the rather neutral absorber
itself. Unfortunately the comparison between different observations,
typically performed at intervals of months to years
(as those discussed here), provides only upper limits to the
intrinsic time scales of  \nhsym\ variations and thus  on the
possible location of the thicker obscuring material (obscuring
``torus" vs. Broad Line Region clouds). The low exposure of the  \swift\ XRT 2006 observations, when the source was in a state similar to the \xmm\ one, did not allow us to establish the $N_{\rm H}$ variability on smaller time scales (i.e. intra-day) of the single observations . An improvement of the
estimates of velocity, distance  and size  from the central X-ray
source of the obscuring material could be obtained only through
monitoring observational campaigns within a few days or weeks and/or through the
search for \nhsym variations within single long
observation.  For what concerns the ionised absorber, as derived from our first order estimate of its distance  from the central black hole (i.e. within $10^{-3}$ pc),  the most likely
location for   this absorber is    much closer in than the stable
and rather neutral one. \\


\section*{Acknowledgements}
We warmly thank T. Yaqoob for the useful discussion  and for helping us 
while fitting  the {\it Mytorus} model  to mimic the  variable absorber. We thank the anonymous referee for many useful suggestions and
comments that significantly improved the paper.
This research has made use of data obtained from the \suzaku\
satellite and data obtained from the High Energy Astrophysics Science
Archive Research Center (HEASARC), provided by NASA's  Goddard Space Flight Center. 
The authors acknowledge financial support from ASI
(grant n. I/088/06/0, COFIS contract and grant n. I/009/10/0). 
VB acknowledge support from the UK STFC research council.

\label{lastpage}

\end{document}